\documentclass[12pt]{article}
\pdfoutput=1
\usepackage{jhep-mod-2}
\usepackage{bm}
\usepackage{amssymb}
\usepackage{hyperref}
\usepackage{setspace}
\usepackage{xcolor}
\def\d{{\mathrm{d}}}

\makeatletter
\begin{document}

\title{
Evading the Trans-Planckian problem with Vaidya spacetimes
}
\author{Ivan Booth$^\dagger$, Bradley Creelman$^\dagger$, Jessica Santiago$^\ddagger$, {\sf and}  Matt~Visser$^\ddagger$}
\affiliation{$^\dagger$ Department of Mathematics and Statistics,
Memorial University of Newfoundland; \\
\null\qquad St. John's, NL A1C 5S7, Canada.}

\affiliation{$^\dagger$ School of Mathematics and Statistics,
Victoria University of Wellington; \\
\null\qquad PO Box 600, Wellington 6140, New Zealand.}
\emailAdd{ibooth@mun.ca, creelmanbjc@gmail.com; 
\\[0pt]\ \ \ \ \ \ \
santi.jsil@gmail.com, matt.visser@sms.vuw.ac.nz}
\abstract{
Hawking radiation, when treated in the ray optics limit, exhibits the unfortunate \emph{trans-Planckian problem} --- a Hawking photon near spatial infinity,  if back-tracked to the immediate vicinity of the horizon is hugely blue-shifted and found to have had trans-Planckian energy. 
(And if back-tracked all the way to the horizon, the photon is formally infinitely blue-shifted, and formally acquires infinite energy.)
Unruh has forcefully argued that this implies that the Hawking flux represents a vacuum instability in the presence of a horizon, and that the Hawking photons are actually emitted from some region exterior to the horizon. 
We seek to make this idea more precise and somewhat explicit by building a purely kinematical model for Hawking evaporation based on two Vaidya spacetimes (outer and inner) joined across  a thin time-like boundary layer. 
The kinematics of this model is already quite rich, and we shall defer consideration of the dynamics for subsequent work. 
In particular we shall present an explicit calculation of the the 4-acceleration of the shell (including the effects of gravity, motion, and the outgoing null flux) and relate this 4-acceleration to the Unruh temperature.

\medskip
\noindent{\sc Keywords\/}:
Hawking radiation, Unruh radiation, trans-Planckian problem, \\
Vaidya spacetime, thin-shell formalism.

\medskip
\noindent
D{\sc{ate}}:  27 September 2018; \LaTeX-ed \today 

P{\sc acs:}  04.70.Dy; 04.70.-s; 04.62.+v; 04.20.Jb
}

\singlespacing
\maketitle
\clearpage
\def\d{{\mathrm{d}}}
\section{Introduction}

The main goal of this article is to develop a purely kinematical model (albeit very much a toy model) that would make it obvious how to evade the so-called ``trans-Planckian'' problem during early and intermediate stages of the Hawking evaporation process~\cite{Jacobson:1999}. Unruh has repeatedly emphasized that  Hawking's original 1973 calculation is a \emph{ray optics} calculation~\cite{explosions}, not a \emph{wave optics} calculation, and that Hawking's 1973 calculation leads to manifest nonsense if you take any specific photon arriving at future null infinity and (in the ray optics approximation) back-track its null geodesic to a region that approaches too close to the horizon~\cite{Unruh:1976, Unruh:2011, Weinfurtner:2013} --- once the back-tracked null geodesic gets too close to the horizon the (locally measured) energy of the photon is gravitationally blue-shifted to ludicrously large energies; easily exceeding the Planck energy, and in fact easily exceeding the total mass-energy of the known universe. So Hawking's \emph{ray optics} calculation \emph{cannot} be the whole story.
Unruh prefers to phrase things roughly along these lines: The quantum vacuum (specifically the Unruh vacuum state) is unstable in the presence of a horizon, and you should look carefully at what escapes to future null infinity, \emph{and what falls into the black hole}.

Indeed, the well-known textbook by Birrell \& Davies~\cite{B&D} actually has a discussion on exactly this point: Birrell \& Davis indicate how to calculate the renormalized stress energy tensor (static approximation, scalar field, no back reaction), and argue that at spatial infinity there is an outgoing positive energy flux, whereas near the horizon there is a \emph{ingoing negative energy flux}. (This of course is how you get around the classical area increase theorem; the classical energy conditions are violated sufficiently close to the horizon~\cite{gvp4, gvp3, gvp-mg, Bardeen:2017}.)

\begin{figure}[!htbp]
\begin{center}
\includegraphics[height=12cm]{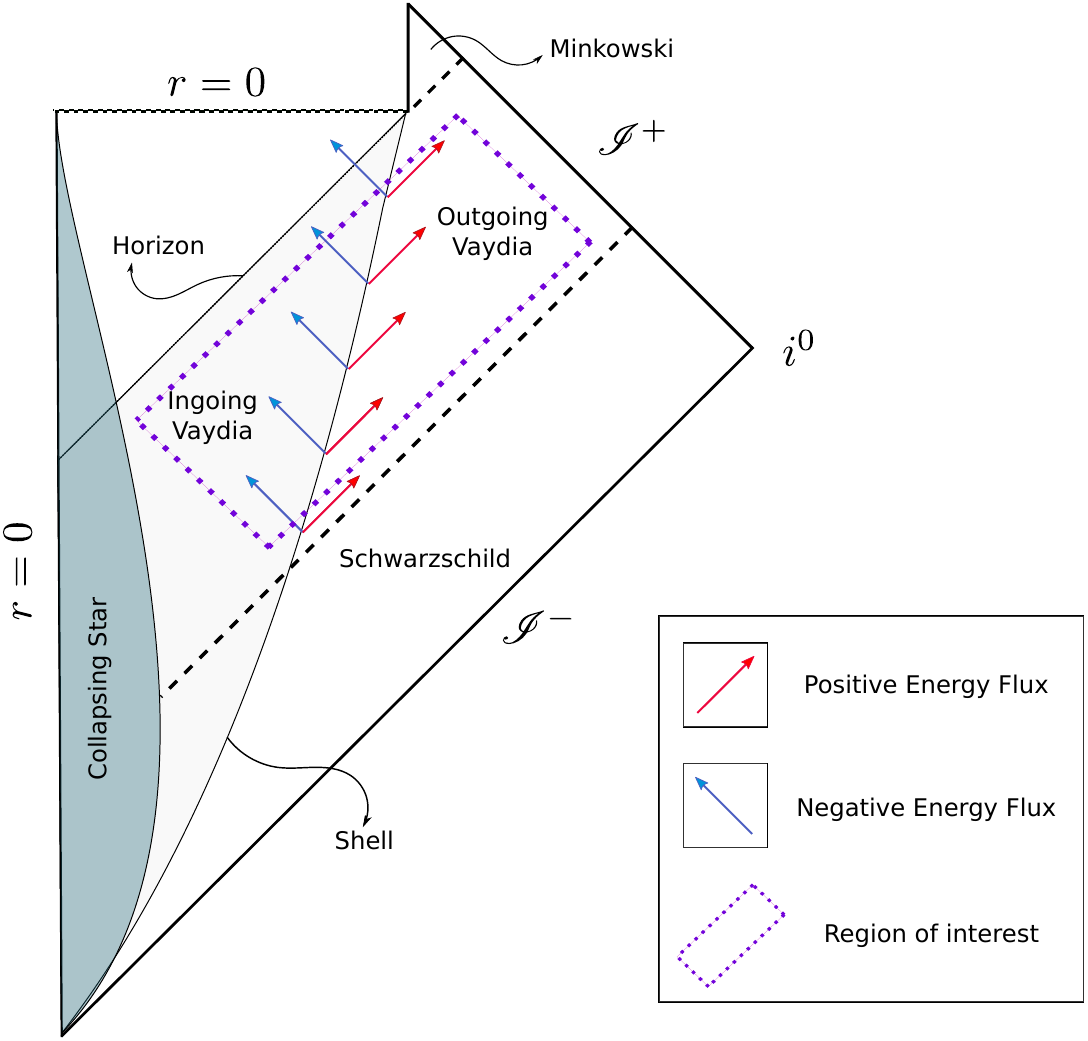}
\caption{Carter--Penrose diagram (one of very many variations on the theme), suitable for matched outgoing and ingoing Vaidya spacetimes as a model for the Hawking evaporation process. This particular model assumes the existence of and pictures an \emph{event} rather than  \emph{apparent} horizon.} 
\label{F:Vaidya}
\end{center}
\end{figure}

For the current article the basic idea we wish to explore is this: Find a toy model (albeit very much simplified) that captures enough of the key behaviour (outgoing positive flux at future null infinity, ingoing negative flux near the horizon), while still remaining reasonably tractable. 
There are several ways of attacking the problem:
\begin{itemize}
\itemsep-3pt
\item Take a general time-dependent spherically symmetric geometry, calculate the Einstein tensor $G_{ab}$, and see if you get anything interesting. (This is likely to be far too flexible a model and therefore unlikely to lead to any significant interesting physical insight.)

\item Take a restricted time-dependent spherically symmetric geometry (say with only one free function in the metric components),  calculate the Einstein tensor $G_{ab}$ and see if you get anything interesting. (This is still likely to be far too flexible a model and still unlikely to lead to significant interesting physical insight.)

\item At large distances, consider a (positive energy flux) outgoing Vaidya ``shining star'' solution~\cite{Vaidya, Vaidya2}; near the horizon consider a (negative energy flux) ingoing Vaidya solution; somehow match them in some intermediate region. This is the approach that we will adopt. Then
there are two options:
\begin{itemize}
\itemsep-3pt
\item One could match across a thick shell; doing this would very much depend on the internal dynamics of the thick shell and  be unlikely to lead to interesting physical insight.
(For instance, one could think of a variant of the Mazur--Mottola thick-shell gravastars~\cite{Mazur:2001, Mazur:2004, Lobo:2005, Visser:2003, Cattoen:2005,  Chirenti:2007}, but with the inner and outer regions now being distinct Vaidya spacetimes.)

\item One could match across a thin shell using the Israel--Lanczos--Sen junction condition formalism~\cite{Israel, Lanczos, Lanczos0, Sen, book}.
(For instance, think of a variant of Mazur--Mottola-like thin-shell gravastars~\cite{MartinMoruno:2011, Lobo:2012, Lobo:2015}, but with the inner and outer regions now being distinct Vaidya spacetimes.)
We shall explore the \emph{kinematics} of such a model in the article below, leaving \emph{dynamics} for future work.\footnote{That is, for the time being we shall only impose the first junction condition, the continuity of the metric, since it is purely kinematical; but for now we shall avoid discussing the second junction condition involving extrinsic curvatures --- the second fundamental forms.} 
(Somewhat related, but distinct in detail, considerations can be found in references~\cite{Boulware, Parentani, Mersini,Baccetti:2017,Vachaspati:2006,Baccetti:2016,Saravani:2012,Alberghi:2001,Barcelo:2007,Barcelo:2009}.)

An advantage of the thin-shell model is that it is as simple as possible (while still capturing key physics). However it also has disadvantages:
\begin{itemize}
\item 
One still needs to decide \emph{where} the transition layer is to be located.
\item
One still needs to make \emph{some} choices regarding the internal dynamics of  the transition layer.
\item 
One still needs to make some choices regarding how the coordinates are set up. 
\item 
In view of the known sparsity of the Hawking flux~\cite{sparse1, sparse2, sparse3}, it should be noted that the Vaidya geometry is a good approximation only on average: a Vaidya-like model necessarily approximates the Hawking flux by a continuum limit. 
\end{itemize}

\item There is neither no real need for, nor advantage in, using generalized Vaidya spacetimes~\cite{Wang:1998qx}.
These all involve extra matter fields, and for our purposes result in more complications without leading to any extra
physical insight. 
\end{itemize}
\item
Thus we choose to focus on two (simple) Vaidya spacetimes matched across a thin shell.
Figure~\ref{F:Vaidya} shows one  of many possible Carter--Penrose diagrams.
\end{itemize}
To set the stage, let us first consider the static approximation, with the Hawking flux treated in the test-field limit (thereby temporarily ignoring back-reaction); 
this idealization closely follows Hawking's original 1973 calculation~\cite{explosions}. Subsequently we shall add back reaction, kinematics, and eventually dynamics. 
This Vaidya-like model can be viewed as a very specific form of black hole ``mimic''; as such this article is complementary to a recent article discussing general phenomenological features of generic black hole ``mimics''~\cite{phenomenology}.

\section{Static approximation --- no back reaction}
Consider first the static approximation in which one ignores back-reaction from the Hawking flux and treats the spacetime geometry as purely Schwarzschild. (After the initial collapse phase, this is  exactly the situation described in Hawking's 1973 calculation; the Hawking flux is a steady test-flux --- its effect on the black hole spacetime is ignored~\cite{explosions}.) Outside the horizon introduce a thin layer located at some (\emph{fixed}, at least for now) radial coordinate $r_s=2G_Nm+\epsilon$, from which we shall assume the Hawking radiation is emitted. (We set $c=1$ and, for now, set $G_N=L_P/m_P$. We find it useful to keep Newton's constant explicit for much of the discussion below.) Conserving energy for the test-flux implies that an equal but opposite ingoing negative energy flux is emitted from the inside of this thin layer, falling into the black hole.

\noindent
Now consider the following quantities:
\begin{itemize}
\itemsep-3pt
\item The (total) gravitational blueshift factor from spatial infinity down to the static thin shell at $r_s$ is:
\begin{equation}
Z = 1+z ={1\over\sqrt{1-{2G_Nm\over r_s}}}  = \sqrt{r_s\over\epsilon} \approx \sqrt{2G_Nm\over\epsilon}.
\end{equation}
\item
A typical Hawking photon has energy $m_P^2/(8\pi m)$ near spatial infinity; but when blue-shifted down to the thin shell this becomes a locally measured  energy of order $[m_P^2/(8\pi m)] \sqrt{2G_Nm/\epsilon}$. For this blue-shifted energy to not exceed the Planck scale (and so avoid the trans-Planckian problem), we require
\begin{equation}
{m_P\over8\pi m} \sqrt{2G_Nm\over\epsilon} \lesssim 1.
\end{equation}
That is
\begin{equation}
\epsilon \gtrsim  {1\over32\pi^2} {G_Nm_P^2\over m} =  {1\over32\pi^2} \;{m_P\over m}\; L_P.
\end{equation}
But this $\epsilon = r_s-2G_Nm$ is a coordinate distance, not a proper distance. 
\item 
The equivalent proper distance measured along any surface of constant-$t$ is
\begin{eqnarray}
\ell = \int_{2G_Nm}^{r_s} {dr\over \sqrt{1-2G_Nm/r}} &\approx&  \sqrt{2G_Nm} \int_{2G_Nm}^{2G_Nm+\epsilon} {dr\over \sqrt{r-2G_Nm}}
\nonumber\\ &&
=  \sqrt{2G_Nm} \left[ 2 \sqrt{r-2G_Nm}\right]_{2G_Nm}^{2G_Nm+\epsilon}.
\end{eqnarray}
That is 
\begin{equation}
\ell \approx \sqrt{8G_Nm\epsilon} \gtrsim {L_P\over2\pi}.
\end{equation}
So, as long as the thin layer is more than a (proper distance) Planck length above the horizon, the trans-Planckian problem is obviated. 

\item This picture has some resemblance to the ``stretched horizon'' (typically invoked by particle physicists)~\cite{stretched} and the ``membrane paradigm'' (typically invoked by classical general relativists)~\cite{membrane,membrane2}.

\item
Even if we choose to work beyond the thin-layer approximation, for some ``thick'' shell, this analysis indicates that as long as the Hawking radiation is emitted from some region more than a (proper distance) Planck length above the horizon, then the trans-Planckian problem is still obviated.
\item
A thin shell held at fixed radial coordinate $r_s=2G_Nm+\epsilon$ undergoes a 4-acceleration of magnitude
\begin{equation}
A = {G_Nm/r_s^2\over\sqrt{1-2G_Nm/r_s}},
\end{equation}
corresponding to a locally determined Unruh temperature
\begin{equation}
T_U = {A\over2\pi} = {G_Nm/r_s^2\over 2\pi\sqrt{1-2G_Nm/r_s}}.
\end{equation}
When redshifted to spatial infinity using the usual Tolman argument this becomes
\begin{equation}
T_{U,\infty} = {A\over2\pi Z} = {G_Nm/a^2\over 2\pi} = T_H \left(2G_Nm\over r_s\right)^2.
\end{equation}
If, as is commonly though not universally advocated, we want the Unruh effect to \emph{quantitatively} explain the Hawking effect, $T_{U,\infty}\approx T_H$, then we would need to assert $r_s\approx 2G_Nm$, or equivalently $\epsilon \ll 2G_Nm$. (Note that the Hawking flux cannot be \emph{exactly} Planckian, at the very least there will be distortion due to: potential violations of adiabaticity, phase space constraints, and greybody factors~\cite{thermality}.) 
\item
Between these two constraints we have
\begin{equation}
{1\over32\pi^2} \;{m_P\over m}\; L_P \lesssim \epsilon \ll 2 \,G_N m. 
\end{equation} 
In terms of proper distance above the horizon this becomes
\begin{equation}
{L_P\over2\pi}  \lesssim \ell \ll  4 \, G_Nm.
\end{equation} 
So, at least in the static approximation, and if you want the Unruh effect to \emph{quantitatively} explain the Hawking effect, then the natural place to put the thin shell is a few (proper) Planck lengths above the horizon.

\item 
There is an alternative that we shall point out but not further explore: Put the shell well above the horizon, say at the unstable photon orbit, $r_s=3G_Nm$, or at the ISCO,  $r_s=6G_Nm$. In this case the thermal flux reaching spatial infinity is given by the modified temperature $T_{U,\infty} = T_H \left(2G_Nm/ r_s\right)^2 \leq T_H$ which is always (by construction) less than the Hawking temperature. This modified temperature is $({4\over9})^{ths}$ of the usual Hawking temperature if the thin shell is placed at the unstable photon orbit, and $({1\over9})^{th}$ of the usual Hawking temperature if it is placed at the ISCO.  That the numerical value of the temperature differs from Hawking's prediction is not entirely unexpected given that one no longer has null curves skimming along and peeling off from the horizon --- one is now interested in null curves emerging from the surface at $r_s$, and the key parameter is the 4-acceleration of that timelike surface. Taking this option destroys the connection between the Hawking temperature and the ``peeling properties'' of near-horizon null geodesics; and in this class of models it is very difficult to see why the Hawking temperature should be universally related to the surface gravity. Some related ideas along these lines are explored in~\cite{Barbado:2011,Barbado:2016}, but we shall not follow this route in the current article.
\end{itemize}
The task now is to (partially and somewhat crudely) include back-reaction effects by making the spacetime geometry time-dependent in an appropriate manner. We shall do this by letting both $r_s$ and $m$ become time dependent, and having the thin shell connect two Vaidya spacetimes as in
Figure~\ref{F:Vaidya}. Much of the preceding analysis (surprisingly much) survives the introduction of this partial back reaction and non-trivial kinematics.

\section{Piecewise Vaidya spacetime}

The Vaidya spacetime (sometimes known as the ``shining star'' spacetime) in its original version adds outgoing null radiation to Schwarzschild spacetime, and can be used as a good model for the exterior geometry of a spherical shining star~\cite{Vaidya,Vaidya2}. We shall consider both outgoing and ingoing Vaidya spacetimes, and combine the two to build a reasonable kinematical model for Hawking radiation.

\subsection{Vaidya spacetime in null coordinates}
Let us work in null coordinates $(w,r,\theta,\phi)$ and write the Vaidya spacetime in the form
\begin{equation}
\d s^2 = -\left(1-{2G_Nm(w)\over r}\right) \d w^2 \mp 2 \d w \d r + r^2(\d\theta^2+\sin^2\theta\;\d\phi^2), 
\end{equation} 
(see for example~\cite{Vaidya, Vaidya2}).
Then, the only non-zero component of the Einstein tensor is
\begin{equation}
G_{ww} =  \mp {2 \,G_N\dot m(w)\over r^2} \, , 
\end{equation}
where the overdot corresponds to a derivative with respect to $w$.
The upper $-$ sign corresponds to outgoing Vaidya spacetime while the lower $+$ sign corresponds to ingoing Vaidya spacetime.

For convenience we rewrite this in the form
\begin{equation}
\d s^2 = -f(w)^2 \left(1-{2G_Nm(w)\over r}\right) \d w^2 \mp 2 f(w) \d w \d r + r^2(\d\theta^2+\sin^2\theta\;\d\phi^2) \, , 
\label{GV}
\end{equation} 
which is equivalent to a coordinate transformation: $w \to \int f(w)\; \d w$ so  $\d w \to f(w) \;\d w$.
Then the non-zero components of the Einstein tensor becomes
\begin{equation}
G_{ww} =  \mp {2 \,G_N f(w) \,\dot m(w)\over r^2}.
\end{equation}

We then match outgoing to ingoing Vaidya spacetimes across a thin shell using the Israel--Lanczos--Sen formalism~\cite{Israel, Lanczos, Lanczos0, Sen, book}.

\subsection{Matching null coordinates outside/inside}

Using the metric in the form (\ref{GV}) there is no loss of generality in using a \emph{common}
coordinate  $w$ for \emph{both} inside and outside. To keep it continuous we introduce
two matching functions $f_\pm(w)$. Then we join the two metrics
\begin{equation}
\d s^2 = -f_\pm(w)^2 \left(1-{2G_Nm_\pm(w)\over r}\right) \d w^2 -\left(  \pm 2 f_\pm(w) \d w \d r \right)
+ r^2(\d\theta^2+\sin^2\theta\;\d\phi^2), \label{metf}
\end{equation} 
across the surface
\begin{equation}
(w, r_s(w), \theta,\phi) \, . 
\end{equation}
In this instance, the metric is written with a  $- (\pm \dots )$ so that the $+$'s and $-$'s in the metric and coordinate functions match up. Then subscript ``$+$'' functions correspond to the outside region and subscript ``$-$'' functions
to the inside.

Thus the (toy) model is completely specified
by the two mass functions $m_\pm(w)$, the two functions $f_\pm(w)$, and the location of the shell $r_s(w)$. 
More precisely it is the ratio $f_+(w)/f_-(w)$, rather than exact functions $f_\pm(w)$, that is physically relevant: Under a 
reparameterization $w\to h(w)$  we can modify both $f_\pm(w)$ though the ratio $f_0(w) = f_+(w)/f_-(w)$ remains fixed.
\subsection{Thin-shell tangent and normal}

With an overdot denoting $d/dw$, on the thin-shell  we have  the (non-normalized) tangent and normal vectors:
\begin{equation}
U^a= (1, \dot r_s(w), 0 ,0)^a ; \qquad\qquad  N_a = (-\dot r_s(w), 1, 0, 0)_a = \nabla_a(r-r_s(w)).
\end{equation}
We now extend these vectors $U^a$ and $N_a$ to the entire manifold, and normalize them
\begin{equation}
u^a = {U^a\over\sqrt{-g_{ab} U^a U^b}} = {U^a\over\|U\|};
 \qquad\qquad
n_a  =  {N_a\over\sqrt{g^{ab}N_a N_b}} = {N_a\over\|N\|}.
\end{equation}
Note that by construction $U^a$ and $N_a$ depend only on $w$, not on $r$. The $r$-dependence in $u^a$ and $n_a$ rises only indirectly, via the normalizing functions.

For future convenience note:
\begin{equation}
g_{ab} =\left[\begin{array}{cc|cc}
-f_\pm(w)^2\; \left(1-{2G_Nm_\pm(w)\over r}\right)&\;\;\mp f_\pm(w)&0&0\\ \mp f_\pm(w)  &0&0&0\\ \hline 0&0&r^2&0\\0&0&0&r^2\sin^2\theta  
\end{array}\right],
\end{equation}
and
\begin{equation}
g^{ab} =\left[\begin{array}{cc|cc}
0&\mp \frac{1}{f_\pm(w)}&0&0\\ \mp \frac{1}{f_\pm(w)} &\;\;\left(1-{2G_Nm_\pm(w)\over r}\right)&0&0\\ \hline 0&0&1\over r^2&0\\0&0&0&1\over r^2\sin^2\theta  
\end{array}\right].
\end{equation}
Thence explicitly
\begin{equation}
U_a = \left(-f_\pm(w)^2\;\left(1-{2G_Nm_\pm(w)\over r}\right)\mp f_\pm(w)\, \dot r_s(w) ,\; \mp f_{\pm} (w)\; ;\; 0, 0 \right),
\end{equation}
and
\begin{equation}
\qquad
N^a = \left(\mp \frac{1}{f_\pm(w)} , \;\left(1-{2G_Nm_\pm(w)\over r}\right) \pm  \frac{\dot r_s}{f_\pm(w)}\;;\; 0, 0 \right).
\end{equation}
The normalizing functions are explicitly given by:
\begin{equation}
\|U\| = \sqrt{-g_{ab} U^a U^b} = \sqrt{f_\pm(w)^2\;(1-2G_Nm_\pm(w)/r)\pm 2 f_\pm \dot r_s(w)} \, , 
\end{equation}
and
\begin{equation}
\| N \| = \sqrt{g^{ab}N_a N_b} = \sqrt{(1-2G_Nm_\pm(w)/r)\pm 2 f_\pm(w)^{-1} \dot r_s(w) } \, ,
\end{equation}
whence
\begin{equation}
\| N \| ={\| U \|\over f_\pm(w)}.
\end{equation}

\subsection{Some technical asides}
We now undertake some technical calculations that will be used later on,  when we calculate the 4-acceleration of the shell.

\subsubsection{The on-shell induced Levi--Civita tensor}

The on-shell metric determinant is
\begin{equation}
\sqrt{-g} = f_\pm(w)\, r^2\, \sin\theta  \, . 
\end{equation}
Further (this will be used in calculating the 4-acceleration)
\begin{equation}
N^a U^b - N^b U^a = (N^w U^r - N^r U^w) \; f_{\pm} \; \varepsilon^{ab}.
\end{equation}
Here $\varepsilon^{ab}$ is the induced Levi--Civita tensor on the $w$-$r$ plane --- specifically $\varepsilon^{ab}$ is an antisymmetric 2-tensor, and in these particular $(w,r,\theta,\phi)$ coordinates we have   $\varepsilon^{wr} =  f_\pm(w)^{-1} = -\varepsilon^{rw}$. 
Then 
\begin{equation}
\label{E:N^U}
N^a U^b - N^b U^a = -\left([1-2G_Nm_\pm/r] \pm 2 f_\pm^{-1} \dot r_s\right) \; f_\pm\;  \varepsilon^{ab} = - \|N\|^2  \; f_\pm\;\varepsilon^{ab}.
\end{equation}

\subsubsection{Exterior derivatives of tangent and normal vectors}

Similarly (this will also be needed when we want to calculate the 4-acceleration of the moving thin shell), it is useful to consider
the exterior derivative
\begin{equation}
\partial_a U_b - \partial_b U_a =-(\partial_w U_r - \partial_r U_w)  \; f_\pm(w)^{-1} \;\varepsilon_{ab},
\end{equation}
where now $\varepsilon_{ab}$ is an antisymmetric 2-form, and in these particular $(w,r,\theta,\phi)$ coordinates we have   $\varepsilon_{wr} = - f_\pm(w) = -\epsilon_{rw}$, so that $\varepsilon^{ab}\varepsilon_{ab}=-2$.
We note that
\begin{equation}
\partial_w U_r - \partial_r U_w = \partial_w (\mp f_\pm) -    \partial_r (-f_\pm^2(1-2G_Nm_\pm/r)\mp f_\pm \dot r_s  ) = 
f_\pm^2 \; {2m_\pm\over r^2} \mp \dot f_\pm.
\end{equation}
That is
\begin{equation}
\label{E:dU}
\partial_a U_b - \partial_b U_a = \left(- f_\pm(w)\; {2G_Nm_\pm(w)\over r^2} \pm {\dot f_\pm(w)\over f_\pm(w)}\right)  \; \epsilon_{ab}.
\end{equation}
Meanwhile $N_a$ is surface forming and so:
\begin{equation}
\partial_a N_b - \partial_b N_a = 0.
\end{equation}

\subsubsection{Normal derivatives}

A more subtle result for the normal derivative (also needed when we want to calculate the 4-acceleration of the moving thin shell), starts from
\begin{equation}
N^a \partial_a = \mp \frac{1}{ f_\pm(w)} \partial_w + \left[\left(1-{2G_Nm_\pm(w)\over r}\right) \pm f_\pm(w)^{-1} \dot r_s \right] \partial_r,
\end{equation}
whence 
\begin{equation}
N^a \partial_a = \mp \frac{1}{f_\pm(w)}\left[\partial_w + \dot r_s \partial_r\right] + \left[\left(1-{2G_Nm_\pm(w)\over r}\right) \pm 2f_\pm(w)^{-1} \dot r_s \right] \partial_r,
\end{equation}
implying
\begin{equation}
\label{E:normal-derivative}
N^a \partial_a = \mp{1\over f_\pm(w)} \; U^a \partial_a + \| N \|^2 \partial_r 
= \mp{1\over f_\pm(w)} \; {d\over d w}  + \| N \|^2 \partial_r.
\end{equation}

\subsection{Constant-$w$ affine null vector}

A particularly obvious and useful constant-$w$ null vector, to be used for defining affine parameters on the radial null geodesics,  is
\begin{equation}
k^a = (0,\pm f_\pm(w)^{-1},0,0);   \qquad k_a = (-1,0,0,0). 
\end{equation}
Here the $\pm$ is chosen to ensure that $k^a$ is future pointing in both regions. 
Now 
\begin{equation}
k^b \nabla_b k^a =  g^{ac} \; k^b\nabla_b k_c,
\end{equation}
and it is easy to see that
\begin{equation}
k^b \nabla_b k_c = k^b (\nabla_b k_c-\nabla_c k_b) =  k^b (\partial_b k_c-\partial_c k_b)  = 0.
\end{equation}
So $k^a = (0,\pm f_\pm^{-1},0,0)$ is the tangent to an \emph{affinely parameterized} null congruence. 

\subsection{Constant-$r$ observer and constant-$r$ normal}

A  ``constant-$r$ observer'' (to be used for defining some notion of ``distance'' to the evolving apparent horizon),  has 4-velocity
\begin{equation}
v^a = {\left(1,0,0,0\right)\over f_\pm \sqrt{1-2m_\pm/r}}; \qquad  v_a = {\left(-f_\pm^2(1-2G_Nm_\pm/r),\mp f_\pm,0,0\right)\over f_\pm \sqrt{1-2G_Nm_\pm/r}}.
\end{equation}
Near spatial infinity (where it makes sense to enforce $f\to1$), this reduces to
\begin{equation}
v^a = {\left(1,0,0,0\right)^a}; \qquad  v_a = {\left(-1,\mp1,0,0\right)_a}.
\end{equation}
In contrast, the non-normalized covariant vector normal to the surfaces of constant $r$ is $(\nabla r)_a=(0,1,0,0)_a$, and the unit normal to the constant $r$ surfaces is
\begin{equation}
\widehat{(\nabla r)}_a = {(0,1,0,0)_a\over\sqrt{1-2G_Nm_\pm/r}}.
\label{E:normal-to-r}
\end{equation}
For completeness we mention that 
\begin{equation}
(\nabla r)^a =(\pm f^{-1}, 1-2G_Nm_\pm/r,0,0)^a
\end{equation}
and
\begin{equation}
\widehat{(\nabla r)}^a = {(\pm f_\pm^{-1},1-2G_Nm_\pm/r,0,0)^a\over\sqrt{1-2G_Nm_\pm/r}}.
\end{equation}

\section{Exterior geometry --- outgoing Hawking radiation}

Let us now consider what happens in the outside region, between the thin shell at $r_s(w)$ and spatial infinity.
It is convenient (and implies no loss of generality) to choose the $w$ coordinate to set $f_+(w)\to1$, and set $m_+(w)\to m(w)$, so that in this exterior region the metric is simply
\begin{equation}
\d s^2 = -\left(1-{2G_Nm(w)\over r}\right) \d w^2 - 2 \d w \d r + r^2(\d\theta^2+\sin^2\theta\;\d\phi^2).
\end{equation} 

\subsection{Blueshift/redshift}
In a dynamic spacetime the general formula for the blueshift/redshift function is~\cite{Stephani}
\begin{equation}
Z  = 1+z ={(k_a V^a)_1\over (k_a V^a)_2}.
\end{equation}
Here we are looking along a null geodesic described by the affine null tangent $k_a$, while $(V^a)_1$ and $(V^a)_2$ are the 4-velocities of the emitter and observer. In the current context
\begin{equation}
Z  = 1+z ={(k_a u^a)\over (k_a v^a)}.
\end{equation}
Here
\begin{equation}
k_a=(1,0,0,0), \quad v_\infty^a = (1,0,0,0), \; \; \mbox{and} \quad u^a = {(1, \dot r_s, 0 ,0)^a \over \| (1, \dot r_s, 0 ,0)^a \|},
\end{equation}
where $v_\infty^a$ is the stationary observer at infinity.

Thus, temporarily reinserting Newton's constant $G_N$ for clarity (and remembering that we are choosing $f(w)\to1$ in the exterior region) the blueshift/redshift from $r=r_s(w)$ to infinity is: 
\begin{equation}
Z_\infty (w)= 1+z ={1 \over \| (1, \dot r_s(w), 0 ,0)^a \|} = {1\over\| U\|}.
\end{equation}
That is
\begin{equation}
Z_\infty (w) = {1\over\sqrt{1-2G_Nm(w)/r_s(w) + 2\dot r_s(w) }}.
\end{equation}

Note how naturally and cleanly this generalizes the static result
\begin{equation}
Z_\infty =1+z ={1\over\sqrt{1-2G_Nm/r_s}};
\end{equation}
there are now (in this non-static evolving situation) contributions both from the gravitational field itself and from the motion of the thin-shell. This computation of $Z(w)$ has significance beyond the thin-shell models considered herein, and would perfectly well apply to a spherically-pulsating ``shining star'' spacetime, as long as the star has a sharp surface at $r_s(w)$ and as long as the stellar exterior is pure outgoing null flux.

It is also worthwhile doing an explicit consistency check by setting $m(w)\to 0$. Then 
\begin{eqnarray}
Z_\infty (w)
&\to& {1\over\sqrt{1+ 2\dot r_s(w) }} \equiv {1\over\sqrt{1+ 2(\d r_s(w)/\d w) }} = {1\over\sqrt{1+ 2{(\d r_s/\d t)\over(\d w/\d t)} }}
= {1\over\sqrt{1+ 2\;{(\d r_s/\d t)\over1 - (\d r_s/\d t)} }}
\nonumber\\
&&
 =\sqrt{1 - (\d r_s/\d t)\over1 + (\d r_s/\d t)}. 
\end{eqnarray}
This (as it should be) is the usual flat-space Doppler shift factor.

\subsection{Evading trans-Planckian physics}
As long as the black hole is ``slowly evolving'' we can use the adiabatic approximation to estimate the average energy of the Hawking photons reaching spatial infinity as
\begin{equation}
E(w) = {m_P^2\over8\pi m(w)}.
\end{equation}
This approximation is valid as long as the surface gravity satisfies $d\kappa/dw \ll \kappa^2$~\cite{Barcelo:2010a,Barcelo:2010b}, that is, as long as $d m(w)/dw \ll m_P/T_P$. That is, this adiabaticity condition is equivalent to the total Hawking luminosity being much less than one Planck power, $L_H \ll m_P/T_P$. There is a similar adiabaticity condition for the validity of Unruh radiation~\cite{barbado}. 

When back-tracked to the thin shell, the Hawking photons will have blueshifted locally measured energy (in the rest frame $u^a$ of the thin shell) given by
\begin{equation}
E_s(w) = {m_P^2 \, Z(w)\over8\pi m(w)}.
\end{equation}
We wish this $E_s(w)$ to be sub-Planckian, that is $E_s(w) \lesssim m_P$, so that
\begin{equation}
{m_P \over8\pi m(w) \sqrt{1-2G_Nm(w)/r_s(w) + 2\dot r_s(w) }} \lesssim 1.
\end{equation}
Writing $r_s(w) = 2G_Nm(w)+\epsilon(w)$, this can be recast as
\begin{equation}
{m_P\over8\pi m(w)} \sqrt{r_s(w)\over\epsilon(w)+ 2 r_s(w) \dot r_s(w)} \lesssim 1.
\end{equation}
That is
\begin{equation}
\epsilon(w)+ 2 r_s(w) \dot r_s(w) \gtrsim {m_P^2\over64\pi^2 m(w)^2} r_s(w).
\label{E:qq1}
\end{equation}

Since for an \emph{evaporating} black hole we must have $\dot r_s(w) < 0$,  this certainly implies
\begin{equation}
\epsilon(w) \gtrsim {m_P^2\over64\pi^2 m(w)^2} r_s(w).
\label{E:qq2}
\end{equation}
Since we want the thin shell to lie outside the Schwarzschild radius, $r_s(w) > 2G_Nm(w) = 2 L_P m(w)/m_P$, this certainly implies
\begin{equation}
\epsilon(w) \gtrsim   {1\over32\pi^2} \;{m_P\over m(w)}\; L_P.
\label{E:qq3}
\end{equation}
This is now a $w$-dependent version of the result we previously obtained in the static approximation.
(While this bound is generally true as long as evaporation overwhelms accretion, $\dot r_s(w)<0$, it is only really a tight bound if both $|\dot r_s| \ll 1$ and $\epsilon(w) \ll 2m(w)$ are small enough.\footnote{The transition from \eqref{E:qq1} to \eqref{E:qq2} is tight only if $\dot r_s|\ll1$. \\
\null\qquad The transition from \eqref{E:qq2} to \eqref{E:qq3} is tight only if $\epsilon(w) \ll 2m(w)$.} Now $|\dot r_s| \ll 1$ will certainly be true during most of the lifetime of the black hole, as long as it is slowly and adiabatically evaporating. Furthermore we shall soon see that $\epsilon(w) \ll 2G_Nm(w)$ will hold if we want the Unruh effect to quantitatively explain the Hawking radiation.)

Let us now estimate the proper distance between the location of the thin shell at $r_s(w)=2G_Nm(w)+\epsilon(w)$, and 
where the apparent horizon ``would have formed'', noting that this  is a ``virtual'' location that is not actually part of the physical spacetime.
To do this, pick some arbitrary but fixed $w_*$ and consider the geometry
\begin{equation}
\d s^2 = -f_+(w_*)^2 \left(1-{2G_Nm_+(w_*)\over r}\right) \d w^2 -\left(2 f_+(w_*) \d w \d r \right)
+ r^2(\d\theta^2+\sin^2\theta\;\d\phi^2). \label{E:frozen}
\end{equation} 
This instantaneously ``freezes'' the external geometry at the moment $w_*$, and then extrapolates it to regions $r<r_s(w_*)$, so that we can say something about where the apparent horizon ``would have formed''. 
Indeed this ``frozen'' geometry is just Schwarzschild geometry in disguise, so all we need to do is to estimate the proper distance between $r_s(w_*)=2G_Nm(w_*)+\epsilon(w_*)$ and $2G_Nm(w_*)$. But this is now standard
\begin{equation}
\ell =  \int_{2G_Nm(w_*)}^{2G)_Nm(w_*)+\epsilon(w_*)} {dr\over\sqrt{1-2G_Nm(w_*)/r}}
\approx\int_{2G_Nm(w_*)}^{2G_Nm(w_*)+\epsilon(w_*)} \sqrt{2G_Nm(w_*) \over r-2G_Nm(w_*)} dr,
\end{equation}
so that
\begin{equation}
\ell \approx \sqrt{8G_Nm(w_*) \epsilon(w_*)}. 
\end{equation}
Since this was calculated for any fixed but arbitrary $w_*$ we see
\begin{equation}
\ell \approx \sqrt{8G_Nm(w) \epsilon(w)}. 
\end{equation}
But then in view of our bound on $\epsilon(w)$ we have
\begin{equation}
\ell \gtrsim {L_P\over2\pi}.
\end{equation}

So even in the presence of back-reaction and an evolving Vaidya spacetime geometry, to avoid trans-Planckian physics we need the Hawking photons to be emitted from a region at least a (proper) Planck length above where the apparent horizon would be expected to form. 
This estimate is subtle --- but there is a good physics reason for the subtlety --- one is making a counter-factual estimate of where the horizon ``would have formed'', an estimate of a hypothetical location that is not part of the actual physical spacetime.

\subsection{From Unruh temperature to Hawking temperature}
\def\L{{\mathcal{L}}}
Let us now calculate $A(w)$ the 4-acceleration of the thin-shell. This 4-acceleration $A(w)$ will be some function of $m(w)$ and $r_s(w)$, and their derivatives. (Calculating the 4-acceleration will be considerably more complicated than in the static approximation.) Spherical symmetry and orthogonality is enough to imply
\begin{equation}
A = n_a  (u^b \nabla_b u^a) = n^a  (u^b \nabla_b u_a) = n^a  (u^b \nabla_b u_a - u^b \nabla_a u_b) 
= n^a u^b (u_{a,b} - u_{b,a}).
\end{equation}

Thence
\begin{eqnarray}
\label{E:4-acceleration}
A &=& {1\over2} (n^a u^b-n^b u^a)\, (u_{a,b} - u_{b,a})
\nonumber\\
&=& {1\over2\|U\|^2\|N\|} (N^a U^b-N^b U^a)\, \left(U_{a,b} - U_{b,a} - {\{U_a \|U\|_{,b} - U_b \|U\|_{,a}]\}\over \|U\|} \right) 
\nonumber\\
&=& {1\over2\|U\|^2\|N\|} (N^a U^b-N^b U^a)\, \left(U_{a,b} - U_{b,a} \right) 
- {{N^a \partial_a} \| U \|\over\|N\| \, \|U\| }.
\end{eqnarray}

But, since in the current situation $f_+\longrightarrow1$ we have $\|U\|=\|N\|$, and in view of equations (\ref{E:N^U}) and (\ref{E:dU}) we have
\begin{equation}
{1\over2}(N^a U^b-N^b U^a)\, \left(U_{a,b} - U_{b,a} \right)  = \|U\|^2\;{2m(w)\over r^2 }.
\end{equation}
Similarly, in view of equation (\ref{E:normal-derivative}) we have
\begin{equation}
{N^a \partial_a} \| U \| = - {d\|U\|\over dw}+ \|U\|^2 \partial_r \|U\| =  - {d\|U\|\over dw}+ \|U\| {m(w)\over r^2}.
\end{equation}
Combining all these results, the 4-acceleration of the thin shell is given by the quite compact and expressive formula
\begin{equation}
A(w) = {1\over \|U\|} {G_Nm(w)\over r_s(w)^2}  +  {1\over \|U\|^2}  {d\|U\|\over dw}
= {1\over \|U\|} \left( {G_Nm(w)\over r_s(w)^2}  + {d\ln \|U\|\over dw}\right). 
\end{equation}

This corresponds to a locally determined Unruh temperature
\begin{equation}
T_U(w) = {A(w)\over2\pi} = {1\over2\pi\|U\|}   \left( {G_Nm(w)\over r_s(w)^2} +  {d\ln \|U\|\over dw}\right).
\end{equation}
When redshifted to spatial infinity (using the previously calculated redshift factor), this becomes
\begin{equation}
T_{U,\infty}(w) = {A(w)\over2\pi Z(w)} = {A(w)\;\|U\|\over2\pi} = {1\over2\pi} \left( {G_Nm(w)\over r_s(w)^2} + {d\ln \|U\|\over dw}\right).
\end{equation}
In terms of the adiabatically evolving Hawking temperature, $T_H(w) = 1/(8\pi G_N m(w))$, where we have set $\hbar=1$ and $c=1$, this is
\begin{equation}
T_{U,\infty}(w) = T_H(w) \;  \left\{ \left(2 G_N m(w)\over r_s(w)\right)^2 +  4 G_N m(w) \; {d\ln \|U\|\over dw}\right\}.
\end{equation}
That is
\begin{equation}
T_{U,\infty}(w) = T_H(w) \;  
\left\{ \left(2 G_Nm(w)\over r_s(w)\right)^2 +  2 G_Nm(w) \; {d\ln \left[ 1-2G_Nm(w)/r_s(w) +2\dot r_s(w) \right]\over dw}\right\}.
\end{equation}

If we want the Unruh effect to \emph{quantitatively} explain the Hawking effect, then we need
\begin{equation}
T_{U,\infty}(w) = {A(w)\over2\pi Z(w)} \approx {1\over 8\pi G_N m(w)}. 
\end{equation}
This is equivalent to asserting
\begin{equation}
r_s(w) \approx 2G_Nm(w); \qquad  G_N m(w) \; {d \|U\|\over\ dw} \ll \|U\|.
\end{equation}
Equivalently
\begin{equation}
r_s(w) \approx 2G_Nm(w); \qquad  G_N m(w) \; {d Z(w)\over\ dw} \ll Z(w).
\end{equation}

So as in the static case, also in this Vaidya context, if we want the Unruh effect of the accelerated thin shell to \emph{quantitatively} explain the Hawking effect, then we need the thin shell to hover just above the apparent horizon --- more precisely, just above where the apparent horizon would otherwise be expected to form --- at least one proper Planck length above the apparent horizon to avoid the trans-Planckian problem. Plus we need the ``slowly evolving'' adiabatic constraint on the evolution of the total redshift $Z(w)$. To obtain these results we only needed to consider the exterior region. 

\section{Interior geometry --- ingoing Hawking radiation}

Given that the ``inner'' geometry is ingoing Vaidya, described by some mass function $m_-(w)$,  can we say anything reasonably explicit about the ingoing (negative energy) Hawking radiation and its impact on the central singularity? Can we say anything reasonably generic regarding the relevant Carter--Penrose diagrams?
Since for current purposes we are interested only in the interior region we can, for the time being, 
set $f_-(w)\to1$ and  $m_-(w)\to m(w)$, so the inner metric takes the form
\begin{equation}
\d s^2 = -\left(1-{2G_Nm(w)\over r}\right) \d w^2 + 2 \d w \d r + r^2(\d\theta^2+\sin^2\theta\;\d\phi^2).
\end{equation} 
Without further loss of generality we have
\begin{equation}
R_{abcd} R^{abcd} =  C_{abcd}C^{abcd} = {48 [G_Nm(w)]^2\over r^6}.
\end{equation}
The orthonormal components of the Weyl tensor are (small) integer multiples of the quantity $G_Nm(w)/r^3$:
\begin{equation}
C_{\hat w\hat r\hat w\hat r} =-2C_{\hat w\hat \theta \hat w\hat \theta} = -2C_{\hat w\hat \phi \hat w\hat \phi} 
= 2C_{\hat r\hat \theta \hat r\hat \theta} = 2 C_{\hat r\hat \phi \hat r\hat \phi} 
= - C_{\hat \theta\hat \phi\hat \theta\hat \phi}  = -{2G_Nm(w)\over r^3}.
\end{equation}
So, the Weyl tensor is completely determined by the quantity $m(w)/r^3$, while the Ricci tensor is completely determined by $\dot m(w)/r^2$. 

\enlargethispage{20pt}
Now recall that the standard endpoints of the Hawking process are a naked singularity, a remnant, or complete evaporation~\cite{B&D}.
\begin{itemize}
\item 
In the current setup, a (permanent) naked singularity would correspond to
\[
\lim_{w\to\infty} m(w) = m_\infty < 0.
\]
(This would be a very unnatural outcome, requiring the Hawking process to ``overshoot'' complete evaporation, and then drive the central mass \emph{negative}. The current framework is not well-adapted to naked singularities.)~\footnote{The only naked singularity you can get in Schwarzschild spacetime is one with negative mass. Similarly for Vaidya, apart from instantaneous
massless shell-focusing singularities at moments of black hole formation or final dispersal (see below),
the only true naked singularities have negative mass. }

\item
In the current setup, a remnant would correspond to either
\[
\lim_{w\to\infty} m(w) = m_\infty > 0,
\]
or at worst a slow asymptotic approach to zero central mass. 
\item 
In the current setup, complete evaporation would correspond to the central mass vanishing at some finite $w_*$:
\[
\lim_{w\to w_*} m(w) =  0.
\]
For $w<w_*$ the geometry is certainly singular at $r=0$; for $w>w_*$ the geometry is certainly regular at $r=0$. Understanding what happens precisely at $w=w_*$ and $r=0$ is more delicate
as we now consider. 
\end{itemize}

Under very mild conditions (the existence of a Puiseaux expansion~\cite{Puiseaux:1850,Puiseaux:1851,Puiseaux1,Puiseaux2}, a condition that is much less  restrictive than a Taylor expansion) one would have
\begin{equation}
m(w) \sim K_m\, (w_*-w)^\gamma \; H(w_*-w).
\end{equation}
Here $H(x)$ is the Heaviside step function, and the critical exponent $\gamma$ controls the behaviour of the final burst (of ingoing negative energy Hawking flux); $K_m$ is some fixed but arbitrary constant.
We impose $\gamma>0$ so that the mass goes to zero at $w=w_*$.

In the immediate vicinity of the final evaporation point, $(w_*,0,\theta,\phi)$, the 
null (causal) structure is determined by $0=-dw^2+2\,dw\,dr = dw(2dr-dw)$, so the outgoing null ray is $ r \sim {1\over2} (w_*-w)$, while the ingoing null ray is given by $dw=0$. So timelike trajectories (suitable for an observer) are of the form
\begin{equation}
r_o(w) \sim K_r\, (w_*-w) \; H(w_*-w); \qquad   K_r \in (1/2,\infty).
\end{equation}
Here $K_r$ is some fixed but arbitrary constant.

\begin{figure}[!htbp]
\begin{center}
\includegraphics[height=8cm]{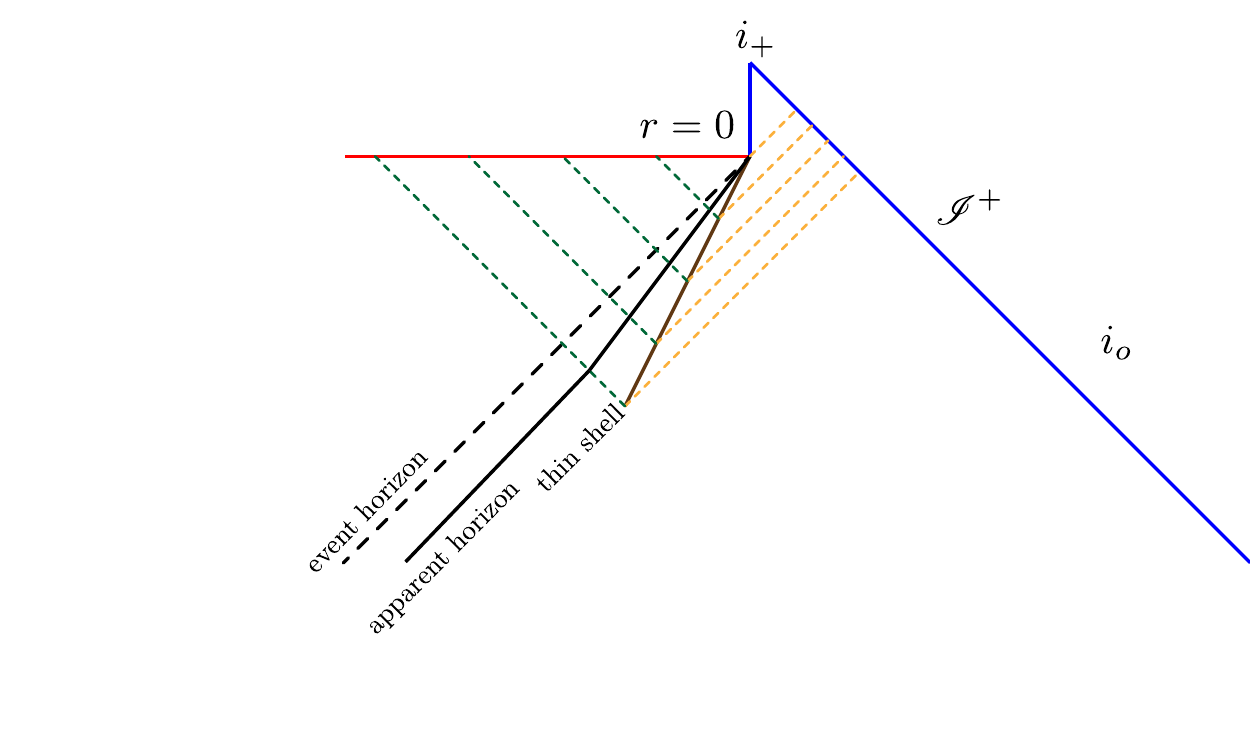}
\caption{Plausible Carter--Penrose diagram focussing on the region near the endpoint of the Hawking evaporation process.
(Under the assumption that a true event horizon forms.)}
\label{F:Carter}
\end{center}
\end{figure}

Therefore, a timelike observer will see orthonormal Weyl components of the form
\begin{equation}
{m(w)\over r_o(w)^3} \sim {K_m\over K_r^3}\;  (w_*-w)^{\gamma-3},
\end{equation}
and orthonormal Ricci components of the form
\begin{equation}
{\dot m(w)\over r_o(w)^2} \sim {K_m\over K_r^2} \; \gamma \;(w_*-w)^{\gamma-3}. 
\end{equation}
\begin{itemize}
\item For $\gamma>3$ the orthonormal components smoothly approach zero.
\item For $\gamma=3$ the orthonormal components at least remain bounded.
\item For $0<\gamma<3$, the orthonormal components blow up.\footnote{Remember that by hypothesis $\gamma>0$.} This corresponds to so-called ``cosmic flashing'', an instantaneous (and not
 particularly troublesome) glimpse of a naked singularity.\footnote{For general spherically symmetric spacetimes (instantaneous) naked massless shell-focusing singularities can also be visible
at moments of black hole formation \cite{Lake:1992zz}.
}
\end{itemize}
Overall, in this framework, complete evaporation seems the most plausible outcome. For all of \emph{these particular models} there will still be an ``information puzzle'' since \emph{by construction} we have assumed the ingoing Vaidya spacetime (with $m_-(w)\neq 0$) to be valid all the way down to $r=0$. (So there will be a true event horizon reaching back from the final evaporation point $w_*$, with concomitant hiding of information behind that event horizon.) To side-step the ``information puzzle'' one would need to modify the ingoing Vaidya spacetime near $r=0$. (See figure~\ref{F:Carter} for a plausible Carter--Penrose diagram presenting the ``standard'' view, where there certainly \emph{is} an ``information puzzle'' due to the \emph{assumed} existence of the event horizon~\cite{observability, existence}.)

\section{Evading the information puzzle}

\begin{figure}[!htbp]
\begin{center}
\includegraphics{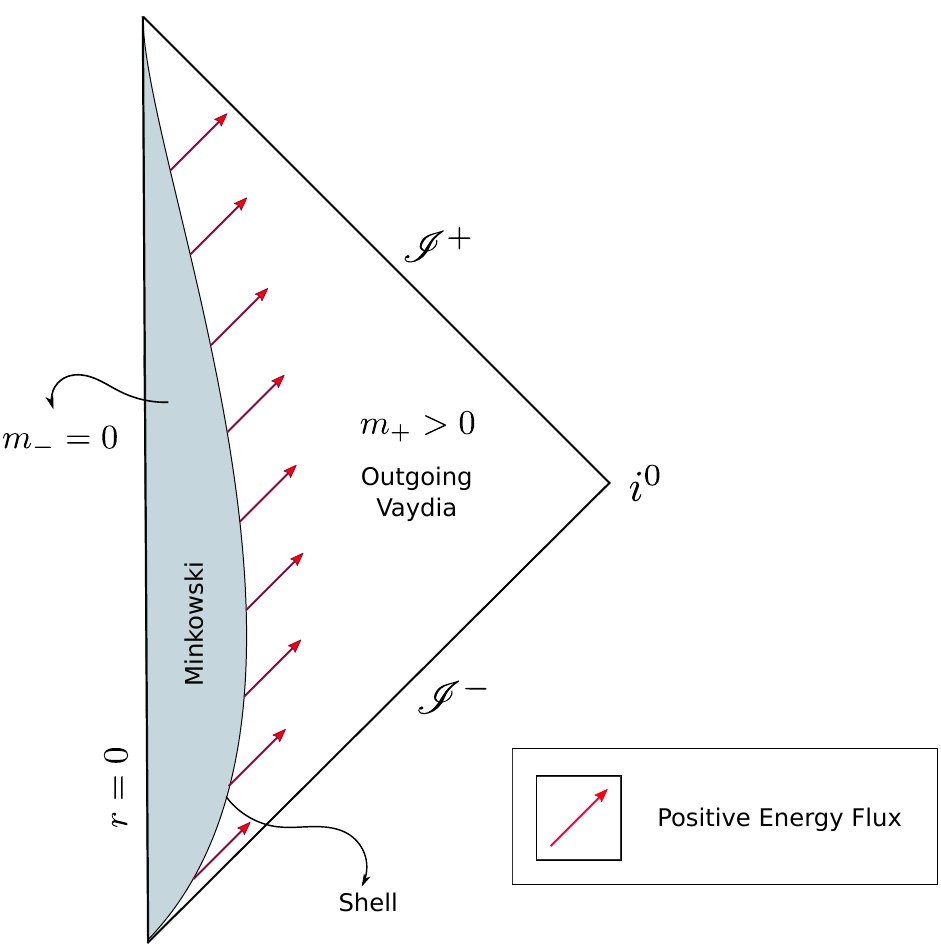}
\caption{Plausible Carter--Penrose diagram for the Hawking evaporation process under the assumption that no event horizon ever forms.}
\label{F:Carter-no-horizon}
\end{center}
\end{figure}

One way of evading the information puzzle is simply to set $m_-(w)\equiv 0$. This corresponds to the interior geometry being completely Minkowski, with all the matter of the ``star'' concentrated on the thin shell, and with an outgoing null flux in the exterior region (figure \ref{F:Carter-no-horizon}).
Models of this type, though undoubtedly somewhat artificial, have previously been considered in the literature, and will be discussed more fully in subsection \ref{ss:empty} below. (See for instance references~\cite{existence,Baccetti:2016,Baccetti:2017}, and the somewhat related~\cite{Saravani:2012,Baccetti:2016-recovery,Saini:2015}. Removing the black hole interior is also a key feature of the ``fuzzball'' proposal~\cite{fuzzball}.) 

The major defect of these particular models is that the inner region is a Minkowski bubble surrounded by the collapsed star --- more reasonable models using black holes with a regular core are also possible. (See for instance references~\cite{Ashtekar:2005,Hayward:2005a,Hayward:2005b,Hayward:2006,Bardeen:2014,Frolov:2016,regular,black-stars}.) 
It is the \emph{assumption} that the mathematically delicate and physically operationally ill-defined concept of \emph{event horizon} survives the introduction of quantum effects that is the source of the information puzzle~\cite{observability,DeLorenzo:2014,Alonso-Serrano:2015,Saini:2015,existence}. 

\section{Models for evaporation scenarios}
\newcommand{\PR}[1]{\ensuremath{\left[#1\right]}}
\newcommand{\PC}[1]{\ensuremath{\left(#1\right)}}
\newcommand{\chav}[1]{\ensuremath{\left\{#1\right\}}}
\newcommand{\ve}[1]{\ensuremath{\langle #1\rangle}}

The formalism we have developed up to this stage is quite generic. We have not yet made any specific choices as to the internal physics of the thin shell as there was already quite a bit that could be said at the purely kinematical level, treating the exterior and interior regions independently. We now  link the exterior and interior regions by enforcing the most basic junction condition --- the continuity of the spacetime metric. This requires (we now set $G_N\to1$)
\begin{equation}
\label{junction}
\left\{-\left(1-{2m_+(w)\over r_s}\right)  - 2 \dot r_s \right\} 
=
\left\{ -f_-(w)^2 \left(1-{2m_-(w)\over r_s}\right)  + 2 f_-(w) \dot r_s \right\}.
\end{equation}
(Here we have, without loss of generality, set $f_+(w)\to 1$.) Now let us look at some interesting possibilities with different mass relations.

\subsection{Mass matching case}

The ``mass matching'' condition, $m_+(\omega) = m_-(\omega) = m(\omega)$,
corresponds to the interior and exterior Vaidya geometries having the same mass function.
If we choose to impose the ``mass matching'' condition, then by (\ref{junction}) either  
$f_-(\omega) = -1$ or
\begin{equation}
f_-(\omega) = 1 + \frac{2\dot{r}_s}{1 -2m(\omega)/r_s} = 1 - \frac{2|\dot{r}_s|}{1 -2m(\omega)/r_s} \label{712}
\end{equation}
where (assuming evaporation) we used $\dot{r}_s <0$.
The $f_-(w)=-1$ option can be safely discarded: By our metric set-up (\ref{metf}) and with $m_+=m_-=m$ this choice 
actually corresponds to attaching the outside metric to a copy of itself, and so represents a radiating white
hole spacetime, rather than an  evaporating black hole.

Now given that the matching surface is timelike and assuming that $\dot{r}_s < 0$ then it follows directly 
from the form of the induced metric that 
\begin{equation}
|\dot{r}_s| < \frac{1}{2}\PC{1 - \frac{2m(\omega)}{r_s}} \, . 
\end{equation}
Hence $f_-(w)$ in  (\ref{712}) is positive. Next defining $r_s = 2m(\omega) + \epsilon$, we have
\begin{eqnarray}
|\dot{r}_s| < \frac{1}{2}\PC{\frac{\epsilon}{2m + \epsilon}} 
\lesssim \frac{\epsilon}{4m(\omega)}
\end{eqnarray}
and so a near-$2m$ transition surface is necessarily slowly evolving. 
This already indicates that the mass matching condition can only be valid for extremely slow shell velocities and, and therefore, for a very slow evaporation. If we now substitute the value of $\epsilon$ that was previously found by analyzing the redshift condition,
\begin{equation}
\epsilon \approx \frac{1}{32 \pi^2} \frac{m_P^2 }{m(\omega)},
\end{equation}
then we obtain:
\begin{eqnarray}
{|\dot{r}_s|}\lesssim \frac{1}{128 \pi^2}\;\; \PC{\frac{m_P}{m(\omega)}}^2.
\end{eqnarray}

In this way, we see that the simple requirement of a timelike matching surface already imposes a condition of very small radial velocity, ensuring that we are dealing with an adiabatic evolution.

\subsection{Non equal masses}
In the completely general case we can extract a quadratic equation for $f_-(\omega)$:
\begin{eqnarray}
f_-^2(\omega) \left(1-{2m_-(\omega)\over r_s}\right)  - 2 f_-(\omega) \dot r_s
-\PR{\left(1-{2m_+(\omega)\over r_s}\right) + 2 \dot r_s} = 0.
\label{fmeq}
\end{eqnarray}
Solving this, we find:
\begin{eqnarray}
f_-(\omega) = \frac{\dot{r}_s \pm \sqrt{\dot{r}_s^2 + \PC{1- 2m_-/r_s}\PR{\PC{1 - 2m_+/r_s} +2\dot{r}_s}}}{(1 - 2m_-/r_s)}.
\end{eqnarray}

\begin{figure}[!htbp]
\center
\includegraphics[scale=0.8]{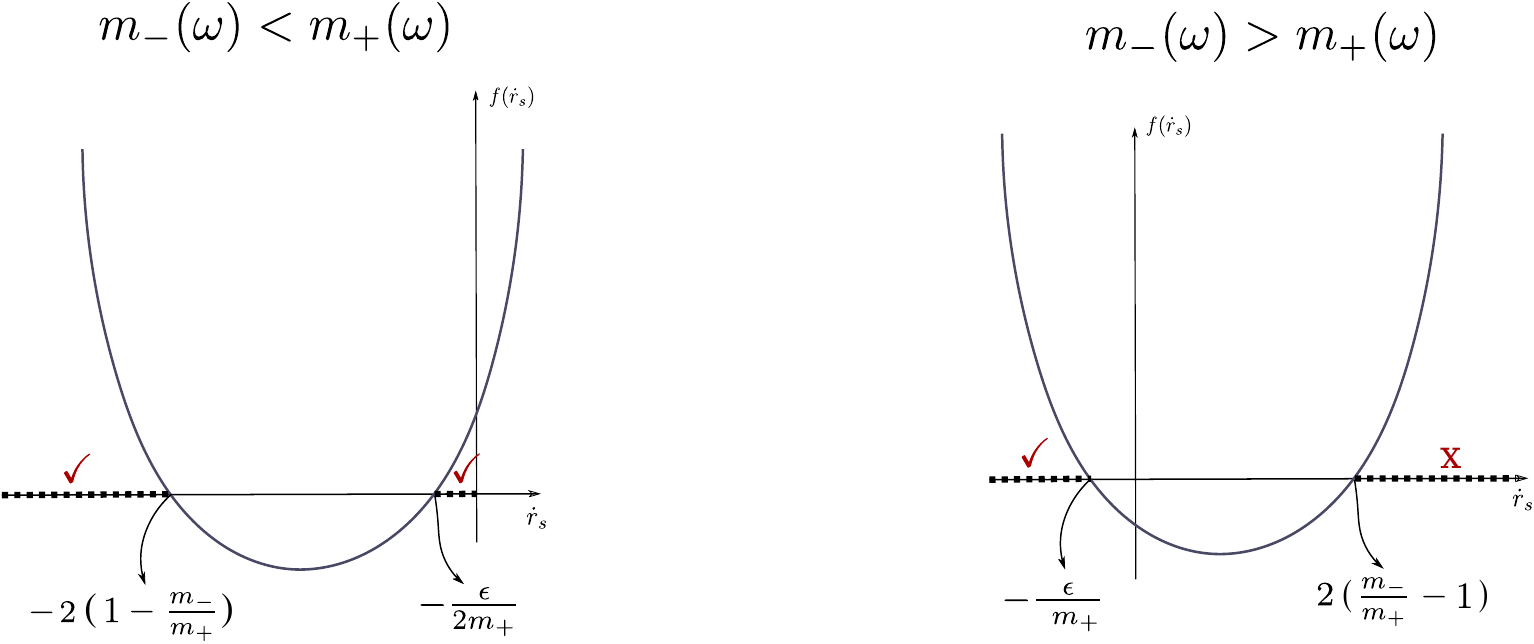}
\caption{Possible scenarios for the radial velocity in the general case.}
\label{F:parabolas}
\end{figure}

We want $f_-(\omega)$ to be real and  so we must have
\begin{equation}
\dot{r}_s^2 + \PC{1- \frac{2m_-}{r_s}}\PR{\PC{1 - \frac{2m_+}{r_s}} +2\dot{r}_s}>0,
\end{equation}
which we rearrange to yield
\begin{equation}
\dot{r}_s^2  + 2\dot{r}_s\PC{1- \frac{2m_-}{r_s}} + \PC{1 - \frac{2m_-}{r_s}}\PC{1 - \frac{2m_+}{r_s}} >0. 
\end{equation}
This places bounds on acceptable values of the model parameters $m_\pm(w)$ and $r_s(w)$. 
Finding the zeros of this quadratic, the edge of the physically acceptable region must satisfy
\begin{eqnarray}
\dot{r}_s &=& 
-\PC{1- \frac{2m_-}{r_s}} \pm \sqrt{\PC{1- \frac{2m_-}{r_s}}^2 - \PC{1- \frac{2m_-}{r_s}}\PC{1- \frac{2m_+}{r_s}}}\\
&=& 
-\PC{1- \frac{2m_-}{r_s}} \pm \sqrt{\frac{2}{r_s}\PC{1- \frac{2m_-}{r_s}}(m_+ - m_-)}.
\end{eqnarray}
Substituting $r_s = 2m_+ + \epsilon$, this becomes
\begin{eqnarray}
\dot{r}_s &=& -1 +\frac{2m_-}{r_s} \pm \sqrt{\frac{2}{r_s^2}
\PC{2m_+ +\epsilon -2m_-}(m_+ - m_-)}.
\end{eqnarray}
If we assume $\epsilon \ll 2\; ||m_+-m_- ||, $ we can approximate this by\footnote{We had already argued $\epsilon \ll 2m_+$ in order for the Unruh effect to be qualitatively linked to the Hawking effect; this $\epsilon \ll 2 ||m_+-m_- ||$ assumption is considerably stronger.}
\begin{eqnarray}
\dot{r}_s 
&\approx & -1 +\frac{2m_-}{r_s} \pm \sqrt{\frac{4(m_+ - m_-)^2}{r_s^2}}
= -1 +\frac{2m_-}{r_s} \pm \frac{2||m_+-m_- ||}{r_s}.
\end{eqnarray}
Then in this case the edges of the physically acceptable region are given by

\begin{equation}
\dot{r}_s^- \approx -1 +\frac{4m_-}{r_s} - \frac{2m_+}{r_s} \approx  -2\left(1 - \frac{m_-}{m_+}\right)
\;\; \mbox{and}\;\;\;\;
\dot{r}_s^+ \approx  -\frac{\epsilon}{2m_+}
\end{equation}
for the $m_+ > m_-\;$ case. Symmetrically
\begin{equation}
\dot{r}_s^- \approx  -\frac{\epsilon}{2m_+}
\;\; \mbox{and}\;\;\;\;
\dot{r}_s^+ \approx  2\left(\frac{m_-}{m_+}-1\right)
\end{equation}
for the $m_+ < m_-\;$ case, which we will explore more closely in section \ref{ss:timematching}.
(For some scenarios, see figure \ref{F:parabolas}.) So, requiring only that $f_-(w)$ has to be real, we already obtain strong restrictions for regions where the model is valid. 

\subsection{The empty-interior massive shell case}
\label{ss:empty}

Consider the case of an ``empty'' interior; when the interior is simply a portion of Minkowski space.  Applying $m_-(\omega)=0$ in equation \eqref{junction}, we obtain:
\begin{equation}
-\left(1-{2m_+(\omega)\over r_s}\right)  - 2 \dot r_s = -f_-(\omega)^2  + 2 f_-(\omega) \dot r_s,
\end{equation}
implying
\begin{equation}
f_-(\omega)^2 - 2f_-(\omega) \dot{r}_s - \PC{1-{2m_+(\omega)\over r_s} + 2 \dot r_s} =0.
\end{equation}
Thence:
\begin{eqnarray}
f_-(\omega) = \dot{r}_s \pm \sqrt{\dot{r}^2_s + \PC{1-{2m_+(\omega)\over r_s} + 2 \dot r_s}}
= \dot{r}_s \pm \sqrt{(1 + \dot{r}_s)^2 -{2m_+(\omega)\over r_s}}.
\end{eqnarray}
We want $f_-(w)$ to be real, so substituting $r_s(\omega) = 2m_+(\omega) + \epsilon(\omega)$, we see
\begin{eqnarray}
(1 + \dot{r}_s)^2 > {2m_+(\omega)\over r_s} \approx 1 - \frac{\epsilon}{2m_+(\omega)}
;\qquad\qquad
|1 + \dot{r}_s| \;\gtrsim\; 1 - \frac{\epsilon}{4m_+(\omega)}.
\end{eqnarray}
As (per assumption) $\dot{r}_s<0$, and taking $|\dot r_s|<1$, so that the evaporation is not ultra-rapid, this implies
\begin{eqnarray}
\label{3}
\dot{r}_s \gtrsim -\frac{\epsilon}{4 m_+(\omega)}; \qquad\qquad
|\dot{r}_s| \lesssim \frac{\epsilon}{4 m_+(\omega)} \ll 1.
\end{eqnarray}
So in this case the velocity of the shell is extremely small, in accordance with the $|\dot{r}_s|$ limitations derived from adiabatic evaporation.

\subsection{The ``time matching" case}
\label{ss:timematching}

If we enforce $f_+(\omega) = f_-(\omega) = 1$, so that coordinate time ``runs at the same rate'' on both sides of the shell, then the matching condition on the shell gives us:
\begin{equation}
m_+(\omega) = m_-(\omega) + 2r_s\dot{r}_s = m_-(w) -2r_s|\dot{r_s}|.
\end{equation}
From this we obtain $\; m_-(\omega) \geqslant m_+(\omega)$. We also want both $m_{\pm}(\omega) > 0$ individually. This gives us:
\begin{eqnarray}
m_- -2r_s|\dot{r_s}| > 0 \qquad \Rightarrow \qquad m_- > 2r_s|\dot{r_s}| . 
\end{eqnarray}
If we substitute $r_s = 2m_+ + \epsilon$,
\begin{equation}
m_- >\; 2(2m_+ + \epsilon)|\dot{r_s}| \;\approx \; 4m_+|\dot{r_s}|,
\end{equation}
giving us an upper limit for the speed of the shell:
	\begin{equation}
|\dot{r_s}| < \frac{m_-(\omega)}{4 m_+(\omega)}.
 \end{equation} 
\begin{figure}
\center
\includegraphics[scale=0.5]{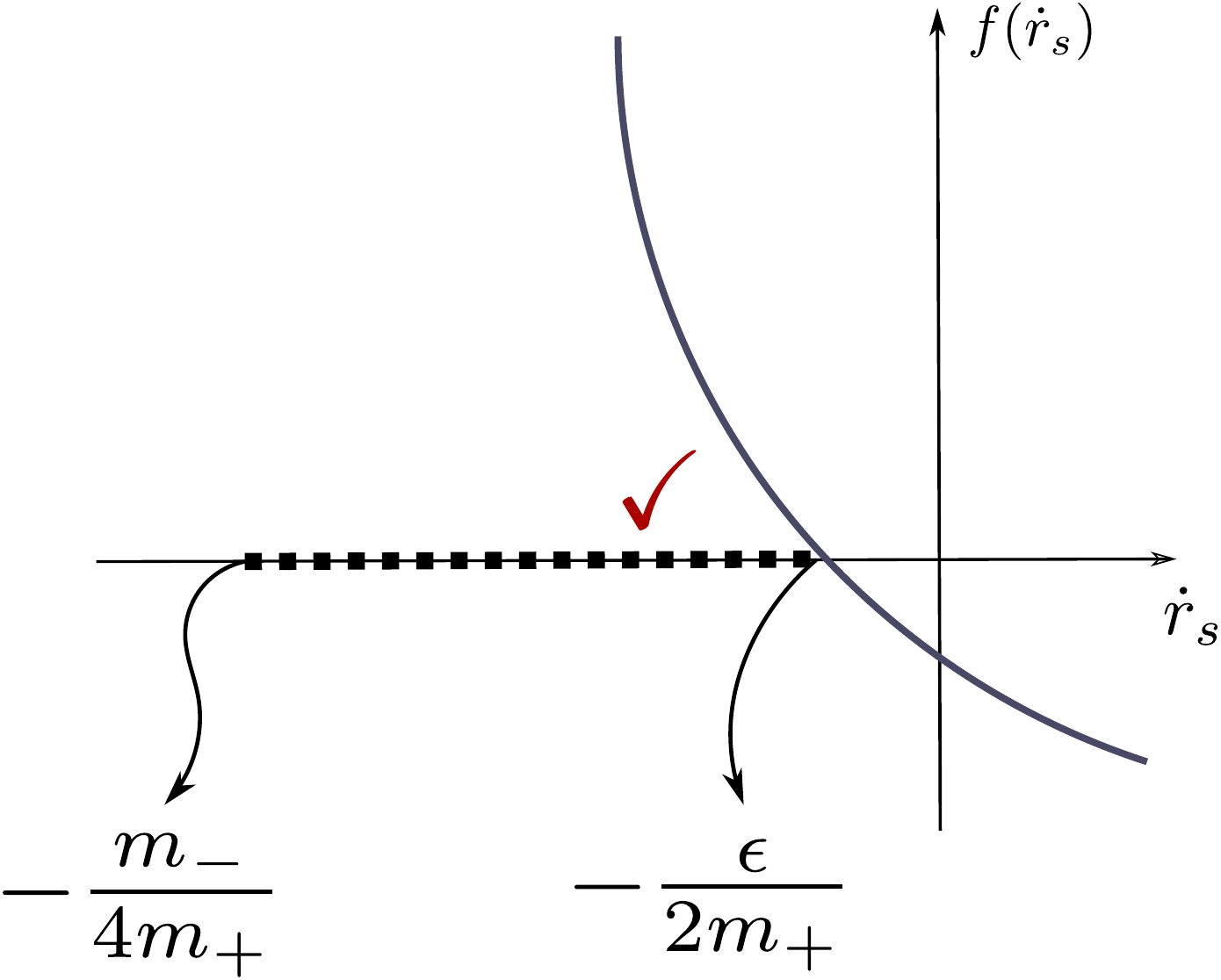}
\caption{The allowed radial velocities for the time matching case.}
\label{time matching}
\end{figure}
As $\;m_- > m_+$, we can use third scenario of figure~\ref{F:parabolas}:
\begin{equation}
\dot{r}_s > -1 +\frac{m_-}{m_+}\qquad\;\hbox{or}\; \qquad \dot{r}_s < -\frac{\epsilon}{2m_+}\; .
\end{equation}
But the first condition gives us positive radial velocities (accretion dominating over the Hawking flux), so we just focus on the second condition. Requiring the interval on figure \ref{time matching} to have a non-zero length we obtain the relatively weak condition $m_-(\omega) > 2\;\epsilon(\omega)$.

The interesting feature here is actually the fact that there is a \emph{minimal} velocity for the evaporation rate of the black hole. Why does this happen? Having $m_- > m_+$ is something that is perhaps a bit unexpected and odd. For that to happen, the thin-shell will have to contain a negative-energy surface density. So the black hole will have to keep evaporating, in order to equilibrate this otherwise unstable situation.

\section{Discussion}

So what have we learned from this exercise? Perhaps the major conclusion to be drawn is that possible scenarios for the final state can be considerably more complex and varied than currently envisaged; and that the sheer number of \emph{a priori} quite reasonable scenarios is quite large --- perhaps unreasonably so. Considerable information can already be extracted at the kinematical level. For instance: (1) Whereas (outgoing) Hawking radiation does not actually seem to need to \emph{cross} the horizon; there are nevertheless good quantitative reasons for believing the Hawking radiation arises from a region near the horizon --- since otherwise there is no good physical reason to connect the surface gravity to the Hawking temperature;
(2) If the interior Vaidya geometry has a nonzero mass $m_-(w)\neq 0$, then some version of the ``information puzzle'' is unavoidable;  being evaded only by the use of an interior ``regular black hole''  or similar construction. 

We have sketched a number of scenarios for the evaporation process, and indicated how very general kinematic considerations can nevertheless lead to interesting constraints on the range of validity of these double-Vaidya thin-shell models. We hope to turn our attention to dynamical issues in future work.

\acknowledgments{
IB is supported by NSERC Discover Grant 2018-04873. BC was supported by
a fellowship from the School of Graduate Studies at Memorial University as well as a stipend from NSERC Discovery Grants 261429-2013.
JS is indirectly supported by the Marsden fund, 
administered by the Royal Society of New~Zealand.
MV is directly supported by the Marsden fund,
administered by the Royal Society of New~Zealand.

A poster presentation outlining some of these results was exhibited by JS at  the conference ``Probing the spacetime fabric: From concepts to phenomenology'' (Trieste, Italy), and at the 2017 Schr\"odinger Institute summer school ``Between Geometry and Relativity''. 
JS also presented these results in a seminar at Nottingham.

\medskip
}

\bigskip
\hrule

\bigskip
\hrule
\bigskip
\end{document}